\documentclass[a4paper]{lipics-v2021}
\usepackage{mathtools}
\usepackage{cite}
\newcommand{\ang}[1]{\langle #1\rangle}

\renewcommand{\ang}[1]{\langle #1\rangle}

\newcommand{\RE}{\mathbb{R}}            

\newcommand{\ST}{\,:\,}                 

\newcommand{\bdOmega}{\partial \kern+1pt \Omega} 

\newcommand{\SP}{\kern+1pt}             
\newcommand{\revFunk}[1]{{}^r \kern-1pt F_{#1}} 
\DeclareMathOperator{\interior}{int}

\usepackage{xcolor}

\title{Ipelets for the Convex Polygonal Geometry}
\titlerunning{Ipelets for the Convex Polygonal Geometry}

\author{Nithin Parepally}{Department of Computer Science, University of Maryland, College Park, USA \and \url{~}}{nparepa@terpmail.umd.edu}{}{}

\author{Ainesh Chatterjee}{Department of Computer Science, University of Maryland, College Park, USA \and \url{~}}{aineshc@terpmail.umd.edu}{}{}

\author{Auguste H. Gezalyan}{Department of Computer Science, University of Maryland, College Park, USA \and \url{~}}{octavo@umd.edu}{https://orcid.org/0000-0002-5704-312X}{}

\author{Hongyang Du}{Department of Computer Science, University of Maryland, College Park, USA \and \url{~}}{hdu1@terpmail.umd.edu}{}{}

\author{Sukrit Mangla}{Department of Computer Science, University of Maryland, College Park, USA \and \url{~}}{smangla@terpmail.umd.edu}{}{}

\author{Kenny Wu}{Department of Computer Science, University of Maryland, College Park, USA \and \url{~}}{wukenny0@gmail.com}{}{}

\author{Sarah Hwang}{Department of Computer Science, University of Maryland, College Park, USA \and \url{~}}{shwang18@terpmail.umd.edu}{}{}

\author{David M. Mount}{Department of Computer Science, University of Maryland, College Park, USA \and \url{https://www.cs.umd.edu/~mount/}}{mount@umd.edu}{https://orcid.org/0000-0002-3290-8932}{}

\authorrunning{Gezalyan, Chatterjee, Du, Hwang, Mangla, Mount, Parepally, Wu}

\Copyright{Auguste H. Gezalyan, Ainesh Chatterjee, Hongyang Du, Sarah Hwang, Sukrit Mangla, David M. Mount, Nithin Parepally, and Kenny Wu}

\relatedversion{}


\date{\today}

\EventEditors{}
\EventNoEds{1}
\EventLongTitle{}
\EventShortTitle{}
\EventAcronym{}
\EventYear{}
\EventDate{}
\EventLocation{}
\EventLogo{}
\SeriesVolume{}
\ArticleNo{}
\hideLIPIcs  
\nolinenumbers 
\begin{document}

\maketitle

\begin{abstract}

There are many structures, both classical and modern, involving convex polygonal geometries whose deeper understanding would be facilitated through interactive visualizations. The Ipe extensible drawing editor, developed by Otfried Cheong, is a widely used software system for generating geometric figures. One of its features is the capability to extend its functionality through programs called Ipelets. In this media submission, we showcase a collection of new Ipelets that construct a variety of geometric objects based on polygonal geometries. These include Macbeath regions, metric balls in the forward and reverse Funk distance, metric balls in the Hilbert metric, polar bodies, the minimum enclosing ball of a point set, and minimum spanning trees in both the Funk and Hilbert metrics. We also include a number of utilities on convex polygons, including union, intersection, subtraction, and Minkowski sum (previously implemented as a CGAL Ipelet). All of our Ipelets are programmed in Lua and are freely available.

\end{abstract}

\section{Structures}
In this section, we describe the geometric structures that our Ipelets compute. Many Ipelets are available that calculate geometric structures such as Poincare disks \cite{IPEletThomasPoincare}, free space diagrams \cite{IPEletRoteFreeSpace}, triangulations\cite{cgalIpe}, circular fillets in polygons\cite{Polyfillet}, graph embeddings\cite{Embeddings}, tangent lines\cite{Tangentlines}, tessellations\cite{Tesselations} and a Voronoi diagrams Ipelet \cite{IPE} (a default Ipelet).

\subsection{Macbeath Regions}

\begin{figure}[htbp]
    \centerline{\includegraphics[scale=0.40]{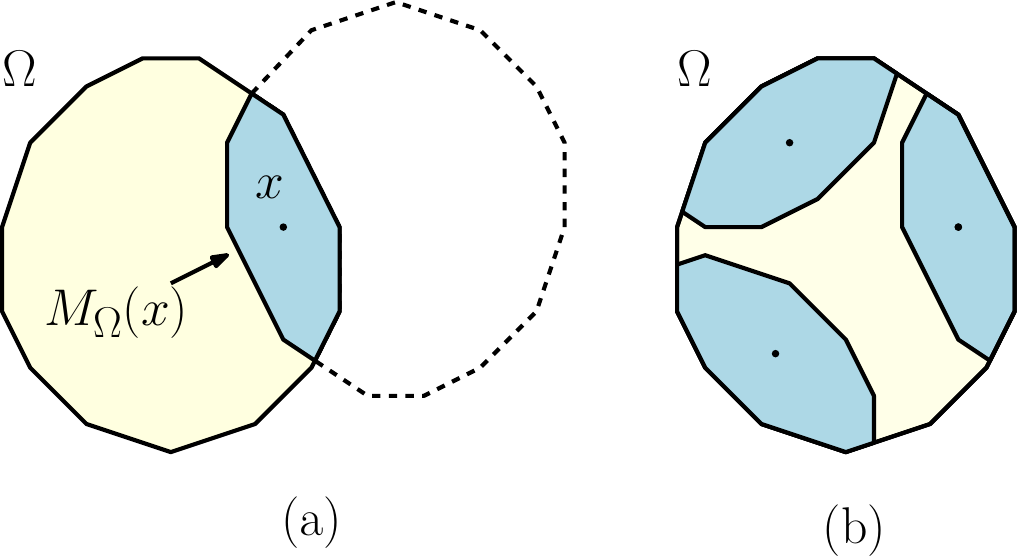}}
    \caption{(a) A Macbeath region and (b) a collection of Macbeath regions} \label{fig:macbeath-2}
\end{figure}

Given a point $x$ in a convex polygon $\Omega$, the Macbeath region\cite{MacbeathMacbeath} around $x$, denoted $M_\Omega(x)$, is a useful tool in convex approximation \cite{arya2012optimal,arya2017optimal,abdelkader2018delone,arya2023economicalpolar}, approximate range searching \cite{arya2006effect,bronnimann1992hard}, smooth distance approximation \cite{abdelkader2023smooth}. Additionally, it has useful relations with cap coverings \cite{barany1988convex,baddeley2007random} and Hilbert balls\cite{abdelkader2018delone}. Macbeath regions around a point are constructed by taking the original polygon and intersecting it with its reflection around the point (see Figure~\ref{fig:macbeath-2}).

\begin{definition}[Macbeath Region] 
Given a convex polygon $\Omega$ in $\RE^d$ and a point $x \in \interior(\Omega)$
\[
    M_{\Omega}(x)
        ~ = ~ x + ((\Omega - x) \cap (x - \Omega)).
\]

\end{definition}

\subsection{Funk and Hilbert Balls}
The Funk metric is an asymmetric metric used in the analysis of Finsler metrics. Its importance for Finsler geometry and differential geometry is the collection of the following three properties: it is non-reversible, complete, and projectively flat \cite{sadeghi2021funk,yu2011new,papadopoulos2019timelike}. The Funk metric also has uses in relation to Hilbert geometries \cite{papadopoulos2014funk, papadopoulos2019timelike}, Weil-Petersson spaces\cite{fujiwara2013geometry}, and hyperbolic billiards \cite{faifman2020funk}. Since it is non-symmetric to define balls, we present Ipelets for both the forward and reverse Funk metric. The Funk and reverse Funk ball around a point $p$ in a polygonal geometry $\Omega$ with $m$ sides is a polygon with $O(m)$ sides. 

Note that in the following definitions, let $\|\cdot\|$ denote the Euclidean norm.

\begin{figure}[htbp]
    \centerline{\includegraphics[scale=0.40]{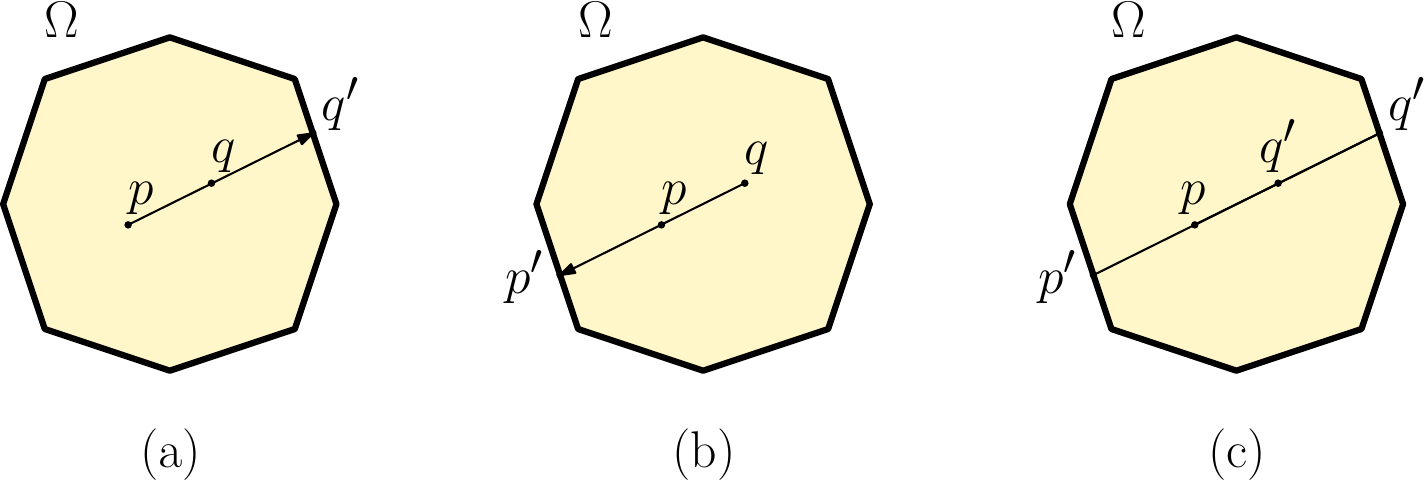}}
    \caption{(a) The Funk metric, (b) the reverse Funk metric, and (c) the Hilbert metric.} \label{fig:FrFH}
\end{figure}

\begin{definition}[(Forward) Funk Metric] 
Given an open convex body $\Omega$ in $\RE^d$ and two distinct points $p, q \in \interior(\Omega)$, let $q'$ be the intersection of the ray $p$ through $q$ with $\bdOmega$ (Figure \ref{fig:FrFH}(a)):
\[
    F_{\Omega}(p,q)
        ~ = ~ \ln \frac{\|p - q'\|}{\|q - q'\|}, ~~\text{and}~~ F_\Omega(p,p)=0.
\]
\end{definition}

\begin{definition}[Reverse Funk Metric] 
Given an open convex body $\Omega$ in $\RE^d$ and two distinct points $p, q \in \interior(\Omega)$, let $p'$ be the intersection of the ray $q$ through $p$ with $\bdOmega$ (see Figure~\ref{fig:FrFH}(b)):
\[
    \revFunk{\Omega}(p,q)
        ~ = ~ \ln \frac{\|p' - q\|}{\|p' - p\|}, ~~\text{and}~~ \revFunk{\Omega(p,p)}=0.
\]
\end{definition}

\begin{figure}[htbp]
    \centerline{\includegraphics[scale=0.40]{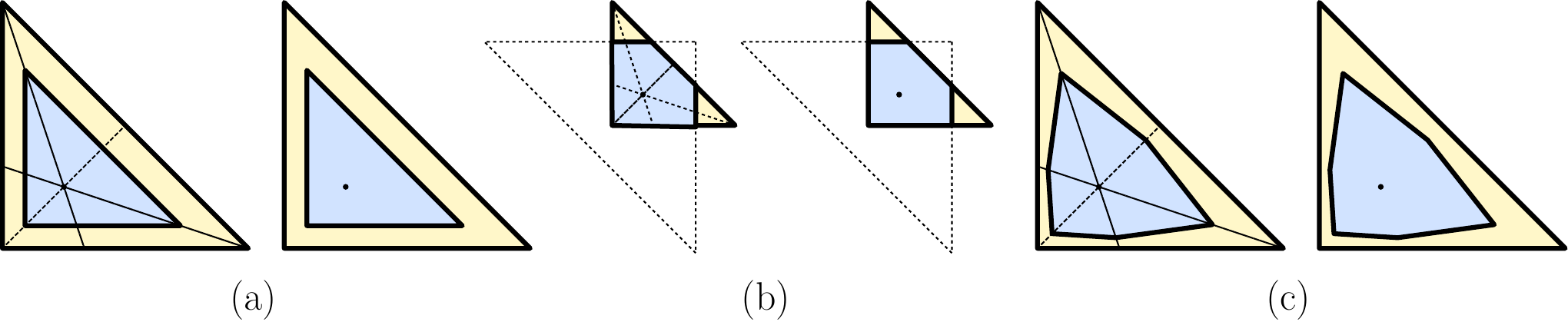}}
    \caption{(a) Funk, (b) reverse Funk, and (c) Hilbert balls (with and without spokes).} \label{fig:HilbertBall}
\end{figure}

The Hilbert polygonal geometry is a generalization of the Cayley-Klein model of hyperbolic geometry to arbitrary convex bodies\cite{colbois2011hilbert} and the average of the forward and reverse Funk metrics. It has become increasingly studied due to its applications in convex approximation \cite{abdelkader2018delone, vernicos2021flag}, clustering \cite{nielsen2019clustering}, and graph embeddings \cite{nielsen2023non} as well as its relation to flags \cite{vernicos2021flag,vernicos2014hilbert}. Several classical algorithms have been developed for the polygonal metric space including Voronoi diagrams \cite{gezalyan2023voronoi,bumpus2023software} and Delaunay triangulations\cite{gezalyan2023delaunay}. The Hilbert ball around a point $p$ in a polygonal geometry $\Omega$ with $m$ sides will be a polygon with $O(m)$ sides\cite{nielsen2017balls}. Like Funk balls, Hilbert balls are constructed with the use of spokes representing the intersection of lines through the center of the ball and the vertices of $\Omega$ with $\bdOmega$ (see Figure~\ref{fig:HilbertBall}). 

\begin{definition}[Hilbert metric] 
Given an open convex body $\Omega$ in $\RE^d$ and two distinct points $p, q \in \Omega$, let $p'$ and $q'$ be the intersection of line $pq$ with $\bdOmega$ such that the points lie in order $\ang{p', p, q, q'}$ (see Figure~\ref{fig:FrFH}(c)) then
\[
    H_{\Omega}(p,q)
        ~ = ~ \frac{1}{2} \ln \left(p',p;q,q'\right)=\frac{1}{2}(F_\Omega(p,q)+\revFunk{\Omega}(p,q)), ~~\text{and}~~ H_{\Omega}(p,p) = 0.
\]
Where $(p',p;q,q')$ is the cross ratio of the four points.
\end{definition}

\subsection{Polar Bodies}
 
\begin{figure}[htbp]
    \centerline{\includegraphics[scale=0.50]{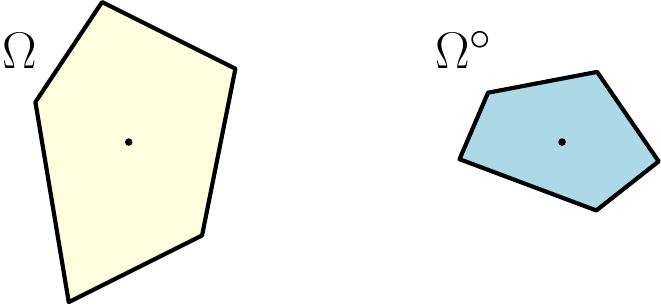}}
    \caption{The polar body.} \label{fig:polar-body}
\end{figure}

The polar body of a convex polygon is a classical duality with modern applications in Minkowski geometry \cite{ghandehari2019minkwoskipolar}, quantum mechanics and information theory \cite{de2022polarquantum,de2023pointillismequantum,de2022symplecticquantum,de2023symplecticquantum}, and convex approximation \cite{naszodi2020approxpolar,arya2022optimalpolar,arya2023economicalpolar}. The key quality of the polar body of a convex polygon is that pointed corners in the primal space become flatter in the dual (see Figure~\ref{fig:polar-body}). 
\begin{definition}[Polar Body]
Given a convex body $\Omega$ containing the origin in $\RE^d$ its polar dual $\Omega^{\circ}$ defined to be:
\[
    \Omega^{\circ}
        ~ = ~ \{ y \in \RE^d \ST \ang{x,y} \leq 1, \forall x \in \Omega \}.
\]
\end{definition}

If $\Omega$ is in $\RE^d$ with vertices $\{(a_i,b_i,\dots)\}_{i=1}^m$, its polar dual $\Omega^{\circ}$ is the polygon defined as the intersection of the half-spaces $\{a_{i}x+b_{i}y +\dots \leq 1\}_{i=1}^m$.

\subsection{Funk and Hilbert Minimum Spanning Trees}

\begin{figure}[htbp]
    \centerline{\includegraphics[scale=0.40]{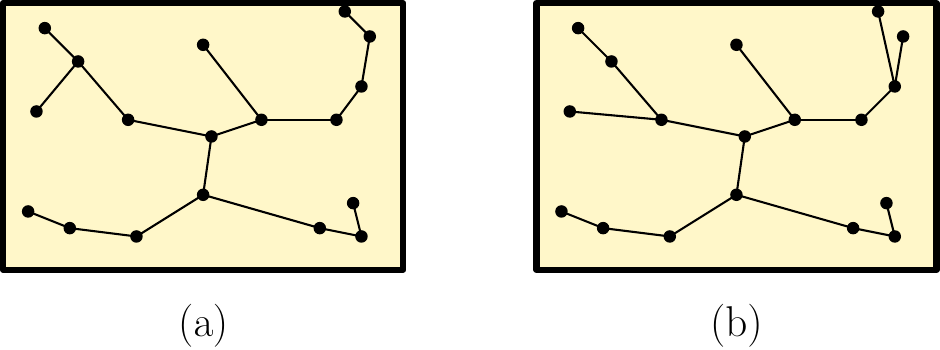}}
    \caption{(a) The Hilbert MST and (b) the Funk MST} \label{fig:HilbertMstFunkMst}
\end{figure}

The minimum spanning tree (MST) of a graph $G$, is the smallest tree contained in $G$ by edge weight that touches every vertex of $G$. Two classical algorithms for the construction of such trees are Kruskal's and Prim's algorithms, which work by incrementally building the MST as a growing forest or tree, respectively. Modern applications include the analysis of brain networks\cite{blomsma2022minimum,habib2022minimum,liu2022minimum}, evolutionary algorithms\cite{bossek2021evolutionary}, data visualization \cite{fischer2023distributed}, and clustering\cite{gagolewski2023clustering}, among many others. The Funk (with respect to the minimum of both the Funk and reverse Funk distance between pairs of points) and Hilbert minimum spanning trees have yet to be studied. An example of each appears below (see Figure~\ref{fig:HilbertMstFunkMst}).

\section{Operations}
In this section we describe the geometric operations that our Ipelets compute.
\subsection{Boolean Operations}
\begin{figure}[htbp]
    \centerline{\includegraphics[scale=0.40]{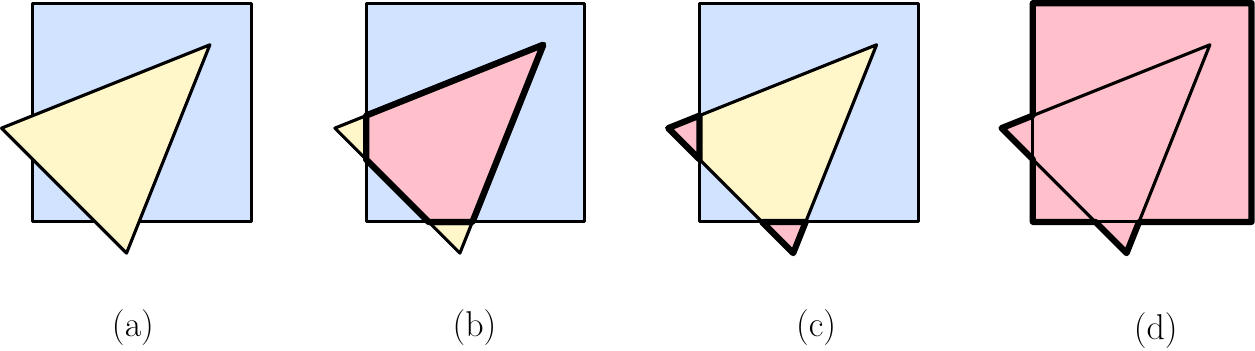}}
    \caption{Boolean operations: (a) original pair, (b) subtraction, (c) union, and (d) intersection.} \label{fig:operations}
\end{figure}

Operations on convex polygons are fundamental to many problems in computational geometry. Boolean operations such as union, subtraction, and intersection are used in geographical information systems\cite{deberg2010book}. We implemented three such operations: polygon subtraction, union, and intersection (See Figure \ref{fig:operations}). 


\subsection{Minkowski Sum}
\begin{figure}[htbp]
    \centerline{\includegraphics[scale=0.40]{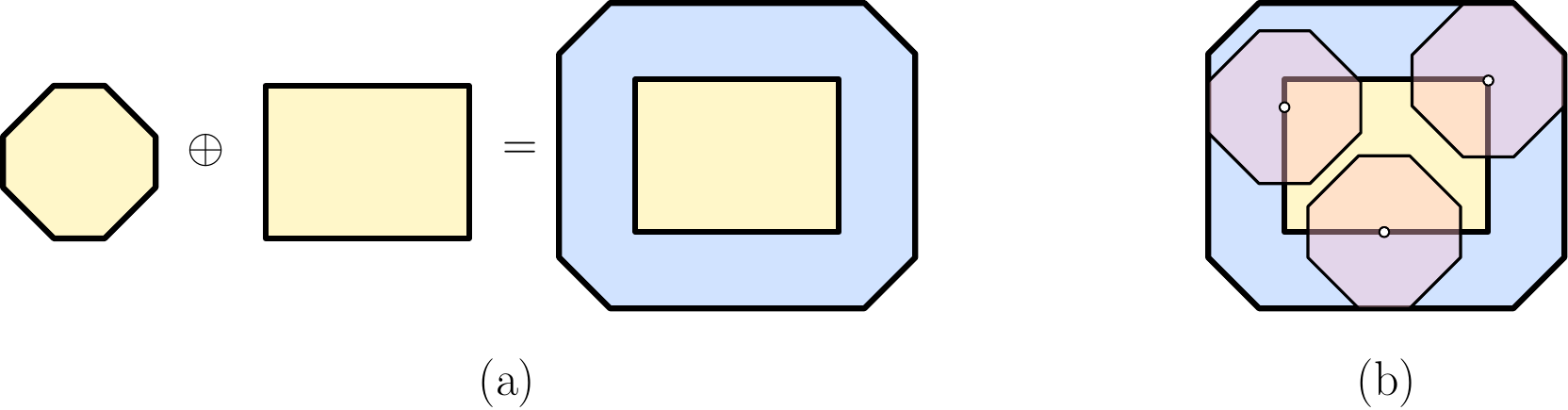}}
    \caption{(a) Minkowski sum of two polygons, (b) collision detection using the two polygons} \label{fig:Minkowski2}
\end{figure}

The Minkowski sum of two polygons is used extensively in many modern fields of research for collision detection \cite{wang2023particlesminkwoski, ruan2019efficientminkwoski, wang2020Minkowskidistance}, solid modeling \cite{diazzi2021convex, sherman2019sound}, and motion planning \cite{guthrie2022closed, graphics2021maximum} others. It is defined as the pairwise addition of all points in both polygons (see Figure \ref{fig:Minkowski2}(a)). The Minkowski sum of two polygons can be thought of as the region traced out by the centroid of one polygon moving along the boundary of the other (see Figure \ref{fig:Minkowski2}(b)). 

\begin{definition}[Minkowski Sum]
Given two polygons $\Omega$ and $\Psi$ the Minkowski sum of the two polygons is defined to be:
\[
    \Omega \oplus \Psi
        ~ = ~ \{ a+b \in \RE^d \ST a \in \Omega, b\in \Psi \}.
\]
\end{definition}

\subsection{Minimum Enclosing Ball}
\begin{figure}[htbp]
    \centerline{\includegraphics[scale=0.40]{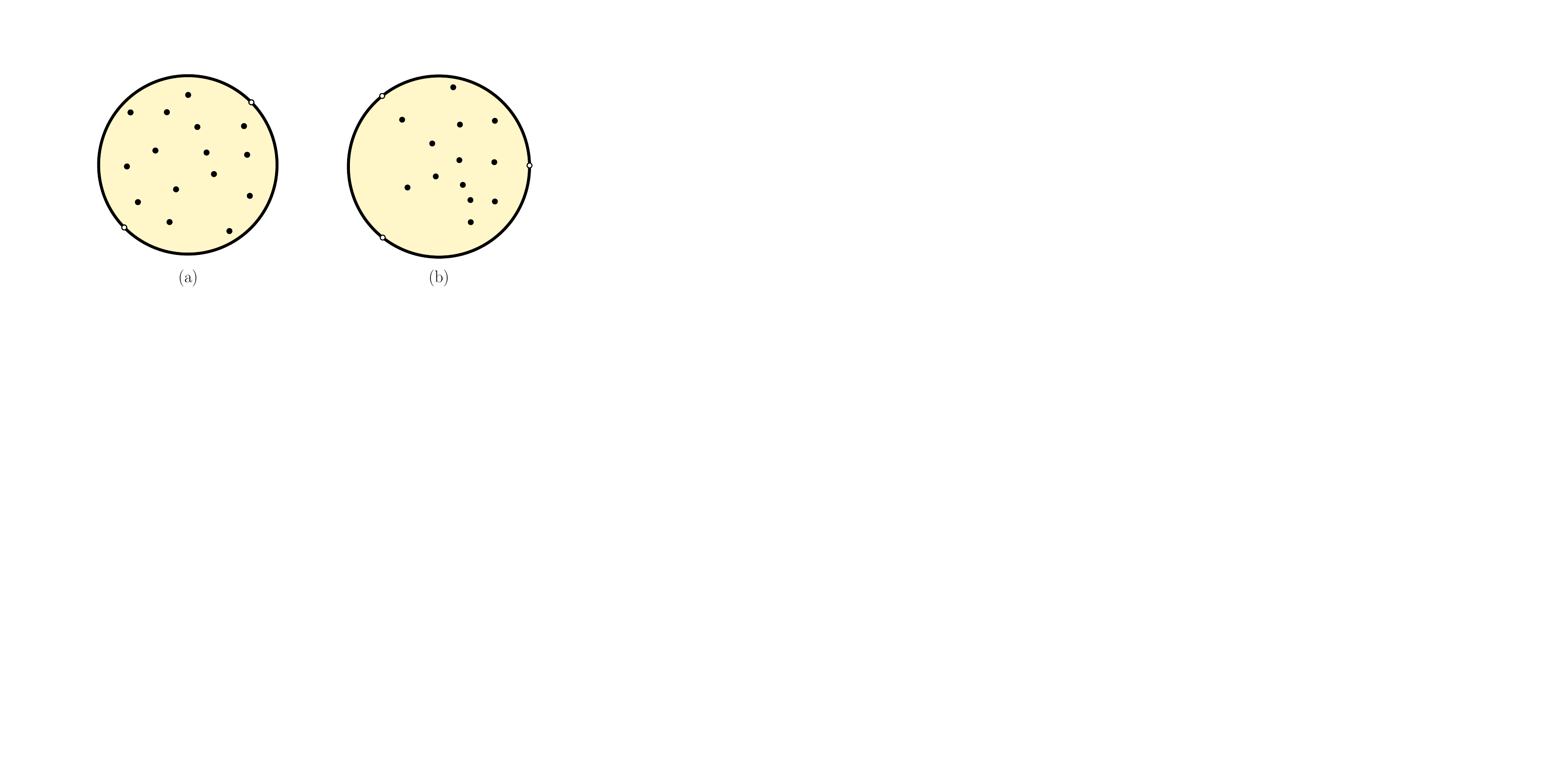}}
    \caption{(a) Minimum enclosing ball defined by two points, and (b) defined by three points} \label{fig:MinimumEnclosing}
\end{figure}

The minimum enclosing ball of a point set $S$ is the smallest ball (by radius) that contains all points in $S$. Some uses of the minimum enclosing ball are vast from genomics\cite{zhou2021scaling}, support vector machines \cite{cervantes2008support}, and forecasting \cite{chandiwana2021twenty}. Note that the minimum enclosing ball of a point-set need not always be defined by three of its points (See Figure \ref{fig:MinimumEnclosing}(a) and (b)).

\section{Installation}
All Ipelets are freely available at \url{https://code.umd.edu/octavo/umd_ipelets/-/tree/main}. To install an Ipelet, download the file and place it in the \texttt{ipelets} subfolder of your Ipe folder.


\bibliography{shortcuts,hilbert}

\end{document}